\documentclass[letterpaper,twocolumn,english,aps,prl,floatfix,superscriptaddress]{revtex4}
\usepackage{float}
\usepackage{amsmath}
\usepackage{amssymb}
\usepackage{graphicx}
\usepackage{esint}

\makeatletter

\pdfpageheight\paperheight
\pdfpagewidth\paperwidth

\@ifundefined{textcolor}{}
{%
 \definecolor{BLACK}{gray}{0}
 \definecolor{WHITE}{gray}{1}
 \definecolor{RED}{rgb}{1,0,0}
 \definecolor{GREEN}{rgb}{0,1,0}
 \definecolor{BLUE}{rgb}{0,0,1}
 \definecolor{CYAN}{cmyk}{1,0,0,0}
 \definecolor{MAGENTA}{cmyk}{0,1,0,0}
 \definecolor{YELLOW}{cmyk}{0,0,1,0}
}

\renewcommand{\[}{\begin{equation}}
\renewcommand{\]}{\end{equation}}

\usepackage{babel}

\usepackage{babel}

\usepackage{babel}

\makeatother

\usepackage{babel}
\begin{document}
\global\long\def\avg#1{\langle#1\rangle}

\global\long\def\p{\prime}

\global\long\def\dg{\dagger}

\global\long\def\ket#1{|#1\rangle}

\global\long\def\bra#1{\langle#1|}

\global\long\def\proj#1#2{|#1\rangle\langle#2|}

\global\long\def\inner#1#2{\langle#1|#2\rangle}

\global\long\def\tr{\mathrm{tr}}

\global\long\def\pd#1#2{\frac{\partial#1}{\partial#2}}

\global\long\def\spd#1#2{\frac{\partial^{2}#1}{\partial#2^{2}}}

\global\long\def\der#1#2{\frac{d#1}{d#2}}

\global\long\def\im{\imath}

\global\long\def\As{{^{\sharp}}\hspace{-1mm}\mathcal{A}}

\global\long\def\Fs{{^{\sharp}}\hspace{-0.7mm}\mathcal{F}}

\global\long\def\Es{{^{\sharp}}\hspace{-0.5mm}\mathcal{E}}

\global\long\def\Fd{{^{\sharp}}\hspace{-0.7mm}\mathcal{F}_{\delta}}

\global\long\def\S{\mathcal{S}}

\global\long\def\A{\mathcal{A}}

\global\long\def\F{\mathcal{F}}

\global\long\def\E{\mathcal{E}}

\global\long\def\O{\mathcal{O}}

\global\long\def\SgF{\S d\F}

\global\long\def\SgEF{\S d\left(\E/\F\right)}

\global\long\def\U{\mathcal{U}}

\global\long\def\V{\mathcal{V}}

\global\long\def\H{\mathbf{H}}

\global\long\def\SO{\Pi_{\S}}

\global\long\def\PO{\hat{\Pi}_{\S}}

\global\long\def\SSH{\tilde{\Pi}_{\S}}

\global\long\def\EO{\Upsilon_{k}}

\global\long\def\ESH{\Omega_{k}}

\global\long\def\HSF{\mathbf{H}_{\S\F}}

\global\long\def\HSEF{\mathbf{H}_{\S\E/\F}}

\global\long\def\HS{\mathbf{H}_{\S}}

\global\long\def\ES{H_{\S}(t)}

\global\long\def\ESo{H_{\S}(0)}

\global\long\def\EgF{H_{\SgF} (t)}

\global\long\def\EgE{H_{\S d\E}(t)}

\global\long\def\EgEF{H_{\SgEF} (t)}

\global\long\def\EF{H_{\F}(t)}

\global\long\def\EFo{H_{\F}(0)}

\global\long\def\ESF{H_{\S\F}(t)}

\global\long\def\ESEF{H_{\S\E/\F}(t)}

\global\long\def\ESSEF{H_{\tilde{\S}\S\E/\F}(t)}

\global\long\def\EEFo{H_{\E/\F}(0)}

\global\long\def\EEF{H_{\E/\F}(t)}

\global\long\def\MI{I\left(\S:\F\right)}

\global\long\def\aMI{\left\langle \MI\right\rangle _{\Fs}}

\global\long\def\BS{\Pi_{\S} }

\global\long\def\PB{\hat{\Pi}_{\S} }

\global\long\def\QD{\mathcal{D}\left(\Pi_{\S}:\F\right)}

\global\long\def\QDp{\mathcal{D}\left(\PB:\F\right)}

\global\long\def\JI{J\left(\Pi_{\S}:\F\right)}

\global\long\def\CI{H\left(\F\left|\Pi_{\S}\right.\right)}

\global\long\def\CIp{H\left(\F\left|\PB\right.\right)}

\global\long\def\CS{\rho_{\F\left|s\right.}}

\global\long\def\CSu{\tilde{\rho}_{\F\left|s\right.}}

\global\long\def\CSp{\rho_{\F\left|\hat{s}\right.}}

\global\long\def\CEF{H_{\F\left|s\right.}}

\global\long\def\CEFp{H_{\F\left|\hat{s}\right.}}

\global\long\def\psiz{\ket{\psi_{\E\left|0\right.\hspace{-0.4mm}}}}

\global\long\def\psio{\ket{\psi_{\E\left|1\right.\hspace{-0.4mm}}}}

\global\long\def\psiinner{\inner{\psi_{\E\left|0\right.\hspace{-0.4mm}}}{\psi_{\E\left|1\right.\hspace{-0.4mm}}}}

\global\long\def\QDz{\boldsymbol{\delta}\left(\S:\F\right)_{\left\{  \sigma_{\S}^{z}\right\}  }}

\global\long\def\NQD{\bar{\boldsymbol{\delta}}\left(\S:\F\right)_{\BS}}

\global\long\def\EFS{H_{\F\left| \BS\right. }(t)}

\global\long\def\EFSM{H_{\F\left| \left\{  \ket m\right\}  \right. }(t)}

\global\long\def\Hol{\chi\left(\Pi_{\S}:\F\right)}

\global\long\def\Holp{\chi\left(\PB:\F\right)}

\global\long\def\ch{\raisebox{0.5ex}{\mbox{\ensuremath{\chi}}}_{\mathrm{Pointer}}}

\global\long\def\rhoS{\rho_{\S}(t)}

\global\long\def\rhoSo{\rho_{\S}(0)}

\global\long\def\rhoSF{\rho_{\S\F} (t)}

\global\long\def\rhoSgEF{\rho_{\SgEF} (t)}

\global\long\def\rhoSgF{\rho_{\SgF} (t)}

\global\long\def\rhoF{\rho_{\F}(t)}

\global\long\def\rhoFp{\rho_{\F}(\pi/2)}

\global\long\def\LE{\Lambda_{\E}(t)}

\global\long\def\LEc{\Lambda_{\E}^{\star}(t)}

\global\long\def\LEij{\Lambda_{\E}^{ij}(t)}

\global\long\def\LF{\Lambda_{\F}(t)}

\global\long\def\LFij{\Lambda_{\F}^{ij} (t)}

\global\long\def\LFc{\Lambda_{\F}^{\star}(t)}

\global\long\def\LEF{\Lambda_{\E/\F} (t)}

\global\long\def\LEFij{\Lambda_{\E/\F}^{ij}(t)}

\global\long\def\LEFc{\Lambda_{\E/\F}^{\star}(t)}

\global\long\def\Lkij{\Lambda_{k}^{ij}(t)}

\global\long\def\Hb{H}

\global\long\def\kE{\kappa_{\E}(t)}

\global\long\def\kEF{\kappa_{\E/\F}(t)}

\global\long\def\kF{\kappa_{\F}(t)}

\global\long\def\ts{t=\pi/2}

\global\long\def\QCB{\bar{\xi}_{\mathrm{QCB}}}

\global\long\def\mc#1{\mathcal{#1}}

\global\long\def\onlinecite#1{\cite{#1}}

\title{Amplification, Redundancy, and the Quantum Chernoff Information}

\author{Michael Zwolak}

\email{mpzwolak@gmail.com}

\address{Department of Physics, Oregon State University, Corvallis, OR 97331}

\author{C. Jess Riedel}

\address{IBM Watson Research Center, Yorktown Heights, NY 10598}

\author{Wojciech H. Zurek}

\address{Theoretical Division, MS-B213, Los Alamos National Laboratory, Los
Alamos, NM 87545}
\begin{abstract}
Amplification was regarded, since the early days of quantum theory,
as a mysterious ingredient that endows quantum microstates with macroscopic
consequences, key to the ``collapse of the wave packet'', and a
way to avoid embarrassing problems exemplified by Schrödinger's cat.
Such a bridge between the quantum microworld and the classical world
of our experience was postulated \emph{ad hoc} in the Copenhagen interpretation.
Quantum Darwinism views amplification as replication, in many copies,
of the information about quantum states. We show that such amplification
is a natural consequence of a broad class of models of decoherence,
including the photon environment we use to obtain most of our information.
This leads to objective reality via the presence of robust and widely
accessible records of selected quantum states. The resulting redundancy
(the number of copies deposited in the environment) follows from the
quantum Chernoff information that quantifies the information transmitted
by a typical elementary subsystem of the environment. 
\end{abstract}
\maketitle
\begin{figure}[H]

\begin{centering}
\includegraphics[width=6cm]{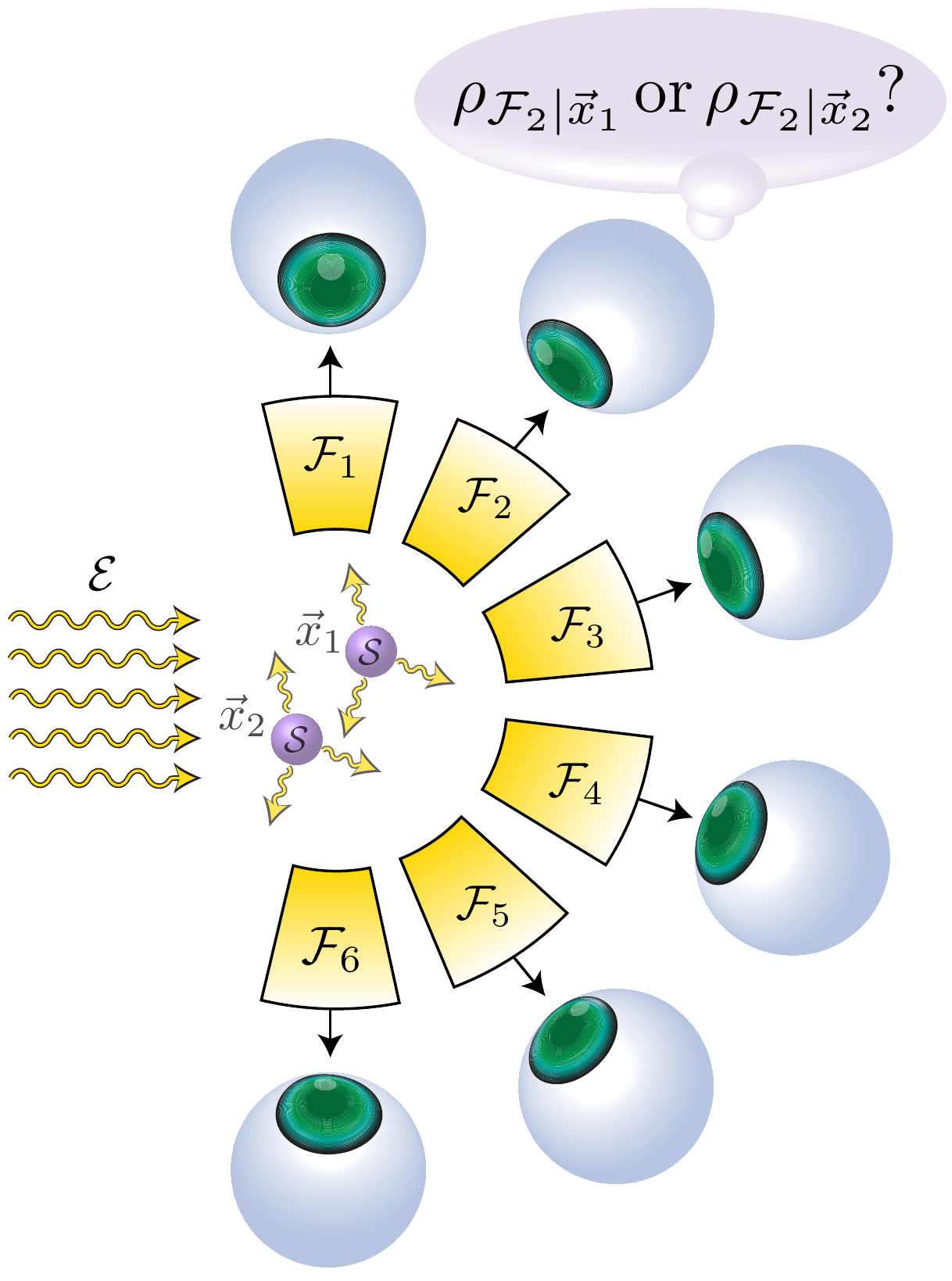} 
\par\end{centering}

\caption{Quantum Darwinism, photons, and the emergence of objective classical
reality.\textbf{ }A quantum system, $\S$, initially in a nonlocal
superposition, is illuminated by the environment, $\E$, composed
of many distinct subsystems (photons) that can be lumped into fragments
$\F$. While $\E$ decoheres $\S$, it acquires many copies of information
about $\S$ that become available to observers who can then independently
infer the state of the preferred (pointer) state of the system without
perturbing $\S$ by direct measurements. This redundant imprinting
of records is responsible for the consensus between observers that
is essential for the emergence of ``objective classical reality''
in our quantum Universe. For the familiar photon environment, the
redundancy (which quantifies amplification) can be enormous: A dust
grain 1 $\mu$m across exposed to sunlight for just 1 $\mu$s will
have its location (to an accuracy of 1 $\mu$m) recorded about $10^{8}$
times in the scattered photons \cite{Riedel10-1}.\label{fig:Schematic}}
\end{figure}

Building on the theory of decoherence \cite{Joos03-1,Zurek03-1,Schlosshauer08-1},
quantum Darwinism is a framework to go beyond the Copenhagen interpretation
and ``bridge'' the quantum-classical divide \cite{Zurek09-1}. It
recognizes that the environment acts as a communication channel for
information about a system of interest, $\S$. Observers acquire information
indirectly by intercepting a fragment $\F$ of the environment $\E$,
such as scattered photons, as happens in everyday life (see Fig. \ref{fig:Schematic}).
This is possible because correlations are created between $\S$ and
$\E$ when they interact. They can be quantified by the quantum mutual
information $\MI=H_{\S}+H_{\F}-H_{\S\F},$ where $H_{\mathcal{A}}=-\tr\rho_{\mathcal{A}}\log_{2}\rho_{\mathcal{A}}$
are the von Neumann entropies. Correlations between elusive quantum
states, when amplified, dependably lead to ``objective classical
reality''. In this Letter, we prove that a broad class of photon
and photonlike environments always amplify information.

The quantum mutual information is naturally divided into classical
and quantum contributions \cite{Zwolak13-1} -- the Holevo quantity
\cite{Kholevo73-1,Nielsen00-1} and quantum discord \cite{Zurek00-1,Ollivier02-1,Henderson01-1},
respectively. Here, we will focus on the Holevo quantity and hence
the information accessible via $\E$ about the system $\S$ -- about
its pointer observable \cite{Zurek81-1} $\PB=\sum_{\hat{s}}\pi_{\hat{s}}\proj{\hat{s}}{\hat{s}}$,
\begin{equation}
\Holp=H\left(\sum_{\hat{s}}p_{\hat{s}}\CSp\right)-\sum_{\hat{s}}p_{\hat{s}}H\left(\CSp\right).\label{eq:Holevo}
\end{equation}
 The Holevo quantity upper bounds the classical information (information
about the pointer states \cite{Ollivier04-1,Zwolak13-1}) transmittable
by a quantum channel (here, the environment), as well as lower bounds
the quantum mutual information, $\MI\ge\Holp$. In this expression,
$\hat{s}=1,\ldots,D_{\S}$ labels the pointer states, $p_{\hat{s}}$
are their probabilities, and $\CSp=\bra{\hat{s}}\rho_{\S\F}\ket{\hat{s}}/p_{\hat{s}}$
are the ``messages'' about $\S$ transmitted by $\F$ -- the fragment
state conditioned on the system's pointer state $\hat{s}$. We will
focus on the case where $\S$ is two dimensional, although the overall
conclusions hold for higher dimensions. 

For information to be objective -- and therefore for the quantum world
to conform to our everyday experience -- many observers should be
able to access it independently \cite{Ollivier04-1,Blume06-1}. The
missing information about $\S$ is quantified by its entropy $H_{\S}$,
and this information must be redundantly proliferated into the world
for it to be objective. In other words, each observer should only
need a small fragment of the environment to retrieve it. This will
allow many observers to independently determine the state of the system,
and reach consensus about it, accounting for the emergence of objective
classical reality in the quantum Universe. More precisely, the number
of fragments of the environment that contain sufficient information
about $\S$ can be deduced starting from the condition 
\begin{equation}
\avg{\Holp}_{\Fd}\cong\left(1-\delta\right)H_{\S},\label{eq:CondRed}
\end{equation}
 where $\avg{\cdot}_{\Fd}$ designates an average over fragments of
size $\Fd$ and $H_{\S}=H(\PB)$, i.e., the entropy of the system
is the entropy of the pointer observable when the system is decohered.
The fragment size $\Fd$ is the number of subsystems of the environment
(e.g., the number of scattered photons or the number of two level
systems) needed for an observer to acquire $\left(1-\delta\right)H_{\S}$
bits of information, on average, about $\S$. The information deficit,
$\delta$, is the information observers can forgo, e.g., observers
may be satisfied with 90\% ($\delta=10^{-1}$) of the missing information.
The number of copies proliferated into the environment defines the
redundancy (the ``gain'', the figure of merit for amplification)
via 
\begin{equation}
R_{\delta}=\frac{\Es}{\Fd},\label{eq:Red}
\end{equation}
 where $\Es$ is the size the environment.

When environments select, but do not perturb, a definite pointer observable
of a system we shall say that $\E$ \emph{purely decoheres} $\S$.
These situations are characterized by the Hamiltonians 
\begin{equation}
\H=\H_{\S}+\PB\sum_{k=1}^{\Es}\EO+\sum_{k=1}^{\Es}\ESH\label{eq:Ham}
\end{equation}
 with $\left[\PB,\HS\right]=0$ and initial states 
\begin{equation}
\rho\left(0\right)=\rho_{\S}\left(0\right)\otimes\left[\bigotimes_{k=1}^{\Es}\rho_{k}\left(0\right)\right],\label{eq:InitState}
\end{equation}
 where $k$ specifies an environment subsystem %
\footnote{In order to have ``complete'' decoherence, $\PB$ and $\EO$ should
not have degenerate eigenvalues (uniformly for all $k$).%
}\cite{Zwolak09-1,Zwolak10-1}. In this scenario, no transitions are
generated between the pointer states $\hat{s}$ (the eigenstates of
$\PB$ \cite{Zurek03-1,Zurek81-1}). The system can still interact
with $\Es$ independent environment subsystems with arbitrary, and
potentially different, interaction operators $\EO$ and self-Hamiltonians
$\ESH$.

The Hamiltonian, Eq.\ \eqref{eq:Ham}, is exact in the case of some
central spin models \cite{Zurek81-1,Zurek03-1} and can be regarded
as a limiting form of one where $\left[\H_{\S},\PB\right]\neq0$ but
the system's evolution (through $\H_{\S}$) occurs slowly on the time
scale it interacts with the environment. Such a condition is broadly
true in our everyday world, as objects are rapidly decohered by collisions
with air molecules and/or photons \cite{Schlosshauer08-1,Joos85-1}.
Moreover, independence of the environment subsystems is satisfied
essentially exactly for the photon environment, and approximately
when the relevant time scales are much faster than the mixing time
of the environment \cite{Riedel12-1}. We assume independence as a
simplification, as it is thought to be approximately necessary for
preserving the redundancy of information. In the following, we will
prove that it is sufficient.

To estimate $R_{\delta}$, we will apply three inequalities and take
$\Fd\to\infty$. The first inequality is Fano's \cite{Nielsen00-1,Cover06-1}
for $D_{\S}=2$, which gives the lower bound 
\begin{equation}
\Holp\ge H_{\S}-H\left(P_{e}\right),\label{eq:Fano}
\end{equation}
 where $P_{e}$, a function of $\F$, is the error probability to
distinguish the conditional states $\CSp$. For practical purposes,
one could easily substitute the right-hand side in Eq.\ \eqref{eq:Fano}
in the definition for redundancy. However, retaining Eq.\ \eqref{eq:CondRed}
in the definition of redundancy will result in the Fano inequality
leading to a lower bound to $R_{\delta}$. The second inequality is
that established in Ref. \cite{Audenaert07-1}, 
\begin{equation}
\tr\left[A^{c}B^{1-c}\right]\ge\tr\left[A+B-\left|A-B\right|\right]/2\label{eq:IneqQCB-1}
\end{equation}
 for two positive operators $A$ and $B$ and $0\le c\le1$. This
inequality was used to prove one side of the quantum Chernoff bound
(QCB) \cite{Audenaert07-1,Audenaert08-1,Nussbaum09-1}, which generalizes
the classical Chernoff bound to sources of independent and identically
distributed (i.i.d.) \emph{quantum} states. Using Eq. \eqref{eq:IneqQCB-1},
we can upper bound the optimal error probability -- from the Helstrom
measurement \cite{Helstrom76-1} -- for distinguishing the $D_{\S}=2$
states generated on the fragment as 
\begin{equation}
P_{e}\le P_{e}^{\star}=p_{1}^{c}p_{2}^{1-c}\prod_{k\in\F}\tr\left[\rho_{k\left|1\right.}^{c}\rho_{k\left|2\right.}^{1-c}\right],\label{eq:UB-1}
\end{equation}
 where $\rho_{k\left|\hat{s}\right.}$ are the subsystem's state conditioned
on the $\hat{s}$ pointer state of $\S$. Here, the conditional subsystem
states are independent, but not identically distributed (i.e., they
are not i.i.d.). Using $P_{e}^{\star}$ in Eq.\ \eqref{eq:Fano}
will further lower bound the accessible information, and thus further
lower bound $R_{\delta}$. The third inequality is $H\left(P_{e}^{\star}\right)\le P_{e}^{\star}/\ln2-P_{e}^{\star}\log_{2}P_{e}^{\star}$
\footnote{This inequality is equivalent to $0\le P_{e}^{\star}+\left(1-P_{e}^{\star}\right)\ln\left(1-P_{e}^{\star}\right)$,
which can be proven by expanding the logarithm and rearranging the
sum to explicitly show that every term is positive.%
}.

We want to determine the relationship between $\Fd$ and $\delta$
that follows from Eq.\ \eqref{eq:CondRed}, which requires averaging
over fragments of the same size. In principle, this could be difficult
if one attempts to optimize the bound in Eq. \eqref{eq:UB-1} by minimizing
over $c$, as $c$ can depend on $\F$. There are important cases
where the optimum $c$ is independent of $\F$, such as the photon
environment below, spin-1/2 environments, and environments with a
pure initial state. Moreover, for the purposes of bounds one does
not have to do any minimization, as one can take any $c$, e.g., $c=1/2$.
Hereon, we will take $c$ as a constant. Averaging Eq. \eqref{eq:UB-1}
over fragments of size $\Fd$, then taking the logarithm and limit
$\Fd\to\infty$ %
\footnote{Implicit in this limit is that the coupling to each environment subsystem
does not decay too rapidly as the environment size grows. In the case
of the photon environment, this condition is satisfied, as there is
a continuous flux of photons that scatter off the system.%
}, one obtains 
\begin{equation}
-\lim_{\Fd\to\infty}\frac{1}{\Fd}\ln\avg{P_{e}}\ge-\ln\avg{\tr\left[\rho_{k\left|1\right.}^{c}\rho_{k\left|2\right.}^{1-c}\right]}_{k\in\E}\equiv\QCB\label{eq:TQCB}
\end{equation}
 We have introduced a ``typical'' Chernoff information $\QCB$,
which is averaged over all subsystems $k$ in $\E$ %
\footnote{Notice that the typical Chernoff information is the quantity of relevance
for amplification, as we will show. In the case of hypothesis testing
by a single observer, however, one would consider $-\avg{\ln\tr\left[\rho_{k\left|1\right.}^{c}\rho_{k\left|2\right.}^{1-c}\right]}_{k\in\E}$,
as we will discuss in a paper in preparation.%
}. Now we need to do the same averaging and limit in order to connect
$\Fd$ and $\delta$. Using Eq.\ \eqref{eq:CondRed} with the three
inequalities, we have 
\[
\delta H_{\S}\le\avg{P_{e}^{\star}/\ln2-P_{e}^{\star}\log_{2}P_{e}^{\star}}_{\Fd}.
\]
 When $\Es\to\infty$, one obtains for the averaging $\avg{P_{e}^{\star}/\ln2-P_{e}^{\star}\log_{2}P_{e}^{\star}}_{\Fd}=g\left(\Fd\right)\exp\left[-\QCB\Fd\right]$,
where $g\left(\Fd\right)$ is a function with the property $\lim_{\Fd\to\infty}\left[\ln g\left(\Fd\right)\right]/\Fd=0$.
Taking the logarithm and the limit $\Fd\to\infty$ yields

\begin{equation}
r\geq\QCB,\label{eq:RUL}
\end{equation}
 where $r=\lim_{\Fd\to\infty,\Es\to\infty}R_{\delta}\ln\left(1/\delta\right)/\Es$
is a measure of the asymptotic efficiency of the amplification. In
the limiting process, the extrinsic scales $\Es$ and $\delta$ have
been removed from $R_{\delta}$, and one is left with $r$, an intrinsic
property of the model. This lower bound immediately establishes that
decoherence processes given by Eqs. \eqref{eq:Ham} and \eqref{eq:InitState}
always redundantly proliferate information.

In addition to a lower bound for $r$, we can also find an upper bound
in many cases. Here, we show the result for i.i.d.~states and $p_{1}=p_{2}$.
This makes use of the upper bound, $\Holp\leq H\left(\left[1-\sqrt{F\left(\rho_{\F\left|1\right.},\rho_{\F\left|2\right.}\right)}\right]/2\right)$,
where $F$ is the fidelity \cite{Roga10-1}. This is further upper
bounded by $H\left(\left[1-2P_{e}\right]/2\right)$, which yields
\footnote{Here we use that for i.i.d. states the QCB yields the exact exponent
in the asymptotic regime.%
} 
\begin{equation}
r\leq2\QCB.\label{eq:RUB}
\end{equation}
 This shows that the Chernoff information is the same as $r$ up to
a factor of $2$. In the examples we have calculated (including the
photon environment below and spin environments), it is the measure
of efficiency asymptotically. In other words, the estimate 
\begin{equation}
R_{\delta}\simeq\Es\frac{\bar{\xi}_{QCB}}{\ln1/\delta},\label{eq:QCBSupp}
\end{equation}
 is exact asymptotically %
\footnote{When $\QCB$ is large, an observer gains a substantial amount of information
by intercepting a very tiny fragment of the environment, potentially
gaining nearly complete information when acquiring a single environment
subsystem. When this occurs, the information deficit, $\delta$, has
to be smaller than some maximum, ensuring that $R_{\delta}/\Es$ is
always less than one.%
}. The close connection between $\delta$ and $P_{e}$, together with
$P_{e}$'s exponential decay, is responsible for the information deficit
appearing only weakly in the redundancy as a logarithm %
\footnote{We note that three inequalities were used, which can give a poor bound
in some cases of interest. However, due to the connection between
$\delta$ and $P_{e}$, (pre)factors will drop out after taking the
logarithm and the asymptotic limit. Thus, the final result can be
tight.%
}. Equations \eqref{eq:RUL} and \eqref{eq:QCBSupp} are the main results
of our Letter. They demonstrate that pure decoherence always gives
rise to redundant information, except for cases of measure zero (e.g.,
when $\rho_{k}\left(0\right)\propto\mathrm{I}$ for all $k$), and
give a practical estimate of the redundancy.

Our work connects the physical processes that amplify information
with the quantum Chernoff bound. The ratio of the number of copies,
$R_{\delta}$, to the number of subsystems, $\Es$, of the environment
is the efficiency of the copying process: An environment subsystem
is imprinted with $\xi_{QCB}/\ln\left(1/\delta\right)$ ``bits''
of information about $\S$. In this sense, $\xi_{QCB}$ is a measure
of the efficiency of the amplification: When Nature consumes $\Es$
environment subsystems -- the ``raw material'' -- then $R_{\delta}\propto\xi_{QCB}\Es$
copies of the system -- the final product -- are proliferated into
the world. Quantum Darwinism, our discussion suggests, can be regarded
as a new kind of communication channel -- an amplification channel:
The same information gets transmitted over and over again, leading
to perception of objective reality.

Now let's consider an example: a photon environment decohering a small
object initially in a spatial superposition $|\psi_{\mathcal{S}}^{0}\rangle\propto|\vec{x}_{1}\rangle+|\vec{x}_{2}\rangle$
through elastic scattering \cite{Joos85-1,Hornberger03-1,Schlosshauer08-1}.
We assume the object is heavy enough that its recoil is negligible,
and that the wavelength of the light is much longer than the object's
extent. This means that the unitary governing the joint evolution
of the system and environment is $\proj{\vec{x}_{1}}{\vec{x}_{1}}\otimes S_{\vec{x}_{1}}+\proj{\vec{x}_{2}}{\vec{x}_{2}}\otimes S_{\vec{x}_{2}}$when
restricted to the relevant two-dimensional subspace of $\ket{\psi_{\S}}$,
so the Hamiltonian is indeed of the form in Eq.\ \eqref{eq:Ham}.
Here, $S_{\vec{x}}$ are the scattering matrices for a single photon
scattering off the object at position $\vec{x}$. For thermally distributed
radiation which originates from a blackbody covering an arbitrary
subset $\mathbb{B}$ of the unit sphere -- the ``sky'' -- $\mathbb{S}$
(as viewed from the object), the redundancy of information, deposited
in the environment, about the position of the object is calculated
in Refs.~\cite{Riedel10-1,Riedel11-1,Korbicz13-1}. The quantum Chernoff
information yields that result (up to a factor $1+\ln(2\ln2)/\ln(\delta)$,
which approaches unity as $\delta\to0$) much more compactly and sheds
light on the significance of the different factors that appear within
the redundancy.

The photon momentum eigenstates are naturally broken into a tensor
product of the magnitude and direction of the momentum. Since the
scattering is elastic and recoilless, a photon's interaction with
the system can only cause mixing between directional eigenstates inside
a subspace of constant energy. This means that the initial thermal
mixedness of the photons does not compete with information acquisition.
Note, of course, that shorter wavelengths are more efficient at distinguishing
between positions of the system, i.e., they have a higher susceptibility.
On the other hand we will see that, compared to the case of illumination
by a point source \cite{Riedel10-1}, the angular spread due to the
finite size of $\mathbb{B}$ will make it more difficult to acquire
information regarding the position of the object.

If we discretize the photon directional states $\vert\hat{p}\rangle$
into bits with small solid angle $\Delta A$, the initial state of
a blackbody photon $k$ is

\[
\rho_{k}\left(0\right)=\int_{0}^{\infty}dp\; P(p)|p\rangle\langle p|\otimes\frac{\Delta A}{A_{\mathbb{B}}}\sum_{\hat{p}\in\mathbb{B}}|\hat{p}\rangle\langle\hat{p}|
\]
 where $A_{\mathbb{B}}$ is the solid angle covered by $\mathbb{B}$.
The distribution of energy (momentum magnitude) eigenstates $\vert p\rangle$
is $P(p)\propto p^{2}/[\mathrm{exp}(pc/k_{B}T)-1]$ for some temperature
$T$. Importantly, all the directional eigenstates $|\hat{p}\rangle$
in the support $\mathbb{B}$ are initially equally likely because
blackbodies are Lambertian radiators \cite{Lambert,Pedrotti93-1}.
This means that the initial state in a fixed-$p$ subspace is in the
block form 
\begin{equation}
\langle p|\rho_{k}\left(0\right)|p\rangle\propto\left[\begin{array}{cc}
\mathbf{I} & \mathbf{0}\\
\mathbf{0} & \mathbf{0}
\end{array}\right]=Q,\label{eq:InitialPhotonState}
\end{equation}
 for projector $Q$ onto the photon directional eigenstates in $\mathbb{B}$.
Since the spatial position of the object is only recorded in the direction,
not energy, of the outgoing photon, the unitary scattering operator
$S_{\vec{x}}$ (conditional on a position $\vec{x}$ of the system)
obeys $S_{\vec{x}}(\vert p\rangle\otimes\vert\hat{p}\rangle)=\vert p\rangle\otimes(S_{\vec{x}}^{p}\vert\hat{p}\rangle)$,
where $S_{\vec{x}}^{p}$ is the operator restricted to fixed-$p$
subspace. With $Q_{p|\vec{x_{i}}}=S_{\vec{x_{i}}}^{p}QS_{\vec{x_{i}}}^{p\dagger}$
for $i=1,2$, the trace in Eq.\ \eqref{eq:TQCB} is proportional
to

\[
\int_{0}^{\infty}dp\; P(p)\mathrm{\, Tr}[Q_{p|\vec{x_{1}}}Q_{p|\vec{x_{2}}}].
\]
 This is independent of $c$ because the $Q_{p|\vec{x_{i}}}$ are
projectors (so $Q_{p|\vec{x_{i}}}^{c}=Q_{p|\vec{x_{i}}}^{1-c}=Q_{p|\vec{x_{i}}}$
for $c\neq0,1$). Thus, this is a case where the optimization over
$c$ can be performed.

We consider $^{\sharp}\mathcal{E}$ photons in a box of volume $V$,
and then take $^{\sharp}\mathcal{E},V\to\infty$ while holding the
number density $^{\sharp}\mathcal{E}/V$ fixed to obtain the correct
radiation flux. In the position basis, the off-diagonal elements of
the density matrix of the object are suppressed by the decoherence
factor $\Gamma=\mathrm{exp}(-t/\tau_{D})$. The decoherence time $\tau_{D}$
is set by \cite{Joos85-1,Hornberger03-1,Schlosshauer08-1,Riedel10-1}
\[
\frac{t}{2\tau_{D}}=\lim_{^{\sharp}\mathcal{E}\to\infty}{}^{\sharp}\mathcal{E}\bigg(1-\frac{\Delta A}{A_{\mathbb{B}}}\mathrm{Re}\int_{0}^{\infty}dp\; P(p)\sum_{\hat{n}\in\mathbb{B}}\langle\hat{n}|S_{\vec{x_{1}}}^{p\dagger}S_{\vec{x_{2}}}^{p}|\hat{n}\rangle\bigg).
\]

Individual photon momentum eigenstates are diffuse in the $V\to\infty$
limit, so $S_{\vec{x}}^{p}$ approaches the identity operator. Ignoring
higher order terms which disappear in this limit, we find for all
$c$ that 
\begin{align}
 & \ln\left[\mathrm{Tr}(\rho_{k|\vec{x_{1}}}^{c}\rho_{k|\vec{x_{2}}}^{1-c})\right]\approx\mathrm{Tr}(\rho_{k|\vec{x_{1}}}^{c}\rho_{k|\vec{x_{2}}}^{1-c})-1\nonumber \\
 & \quad=\frac{\Delta A}{A_{\mathbb{B}}}\int_{0}^{\infty}dp\; P(p)\sum_{\hat{n}\in\mathbb{B}}\sum_{\hat{m}\in\mathbb{B}}|\langle\hat{n}|S_{\vec{x_{1}}}^{p\dagger}S_{\vec{x_{2}}}^{p}|\hat{m}\rangle|^{2}-1\nonumber \\
 & \quad=\frac{\Delta A}{A_{\mathbb{B}}}\int_{0}^{\infty}dp\; P(p)\sum_{\hat{n}\in\mathbb{B}}\sum_{\hat{m}\in\mathbb{B}}|\langle\hat{n}|(S_{\vec{x_{1}}}^{p\dagger}S_{\vec{x_{2}}}^{p}-I)|\hat{m}\rangle|^{2}\nonumber \\
 & \quad\quad\quad-2\bigg(1-\frac{\Delta A}{A_{\mathbb{B}}}\mathrm{Re}\int_{0}^{\infty}dp\; P(p)\sum_{\hat{n}\in\mathbb{B}}\langle\hat{n}|S_{\vec{x_{1}}}^{p\dagger}S_{\vec{x_{2}}}^{p}|\hat{n}\rangle\bigg)\nonumber \\
 & \quad\to-\frac{\alpha}{^{\sharp}\mathcal{E}}\frac{t}{\tau_{D}},\label{eq:QCBPhotonSupp}
\end{align}
 where 
\begin{align}
\alpha=\frac{\int_{0}^{\infty}dp\; P(p)\int_{\mathbb{B}}d\hat{n}\int_{\mathbb{S}\backslash\mathbb{B}}d\hat{m}|\langle\hat{n}|(S_{\vec{x_{1}}}^{p\dagger}S_{\vec{x_{2}}}^{p}-I)|\hat{m}\rangle|^{2}}{\int_{0}^{\infty}dp\; P(p)\int_{\mathbb{B}}d\hat{n}\int_{\mathbb{S}\phantom{\backslash\mathbb{B}}}d\hat{m}|\langle\hat{n}|(S_{\vec{x_{1}}}^{p\dagger}S_{\vec{x_{2}}}^{p}-I)|\hat{m}\rangle|^{2}},
\end{align}
 is the so-called \emph{receptivity} of the environment to making
records about the system \cite{Riedel11-1}. Its form guarantees that
$0\le\alpha\le1$. We have made use of the definitions of $\tau_{D}$
and $\rho_{k|\vec{x}_{i}}=S_{\vec{x_{i}}}\rho_{k}(0)S_{\vec{x_{i}}}^{\dagger}$,
the completeness relations $\mathrm{I}=\sum_{\hat{m}\in\mathbb{S}}|\hat{m}\rangle\langle\hat{m}|$,
and that $\sum_{\hat{n}\in\mathbb{B}}=\sum_{\hat{n}\in\mathbb{S}}-\sum_{\hat{n}\in\mathbb{S}\backslash\mathbb{B}}$.
($\mathbb{S}\backslash\mathbb{B}$ is the complement set of $\mathbb{B}$
inside $\mathbb{S}$.)

Plugging Eq. \eqref{eq:QCBPhotonSupp} into Eq. \eqref{eq:QCBSupp},
we obtain 
\[
R_{\delta}\simeq\frac{\alpha t/\tau_{D}}{\ln1/\delta},
\]
 which, as $\delta\to0$, is Eqs.~(24) and (25) from Ref.~\cite{Riedel11-1}.
Redundant information is thus generated at a rate of $\alpha/\left(\tau_{D}\ln\left(1/\delta\right)\right)$.
The factors involved signify three essential ingredients of information:
$\ln\left(1/\delta\right)$ reflects the accuracy of the information
desired by an observer; $\tau_{D}$ represents that the environment
and system have interacted which simultaneously decoheres the system
and transfers information; $\alpha$ is how receptive the environment
is to acquiring information. The \textit{redundancy rate} thus has
a remarkably simple and transparent form when evaluated using the
quantum Chernoff information.

\emph{Conclusions.} \textendash{} We demonstrated how processes that
are ubiquitous in the natural world, such as photon illumination,
amplify selected information about quantum systems. Photon and photonlike
environments give rise to the redundant proliferation of information
regarding pointer states -- they are the mechanism by which one original
becomes many. Information can then be accessed simultaneously and
independently by many observers. Objective, classical reality appears
as a consequence. The ``typical'' quantum Chernoff information,
$\QCB$, quantifies the efficiency of the amplification, which is
strictly positive except for measure zero scenarios. The resultant
amplification is huge, as it is linear in the environment size, $\Es\QCB/\ln\left(1/\delta\right)$.
The information disseminated through the environment resides in the
states of its individual subsystems. They allow one to acquire the
information about the pointer states the ``systems of interest''
indirectly, via the fragments of $\E$. This amplification and proliferation
of selected information results in the emergence of (our perception
of) the classical world. The interplay between information available
locally from the environment and its complement (quantified by quantum
discord) explains the origins of objective reality in the quantum
Universe \cite{Zwolak13-1,Streltsov13-1,Brandao13-1} and helps delineate
the quantum-classical border. 
\begin{acknowledgments}
We would like to thank Jon Yard for helpful discussions. We would
also like to thank the Center for Integrated Quantum Science and Technology
(IQST), Universität Ulm, where part of this work was carried out.
This research was supported in part by the U.S. Department of Energy
through the LANL/LDRD Program and, in part, by the John Templeton
Foundation and by FQXi. 
\end{acknowledgments}

\end{document}